\documentclass{frontiersSCNS} 

\graphicspath{{./}{./figs/}}
\usepackage{url,lineno}
\usepackage{xcolor} 
\usepackage[urlcolor=blue]{hyperref}      
\hypersetup{                              
    colorlinks = true,                    
    citecolor = {blue},
    linkcolor = {green},
    urlcolor  = {purple},
           }

\copyrightyear{}
\pubyear{}

\def\journal{Robotics and AI}
\def\DOI{}
\def\articleType{Research Article}
\def\keyFont{\fontsize{8}{11}\helveticabold }
\def\firstAuthorLast{S\'andor {et~al.}} 
\def\Authors{Bulcs\'u S\'andor\,$^{1,*}$, Tim Jahn\,$^{1}$, Laura Martin\,$^{1}$ and Claudius Gros\,$^1$}
\def\Address{
$^{1}$Institute for Theoretical Physics, Goethe University Frankfurt, 
	Frankfurt am Main, Germany}
\def\corrAuthor{Bulcs\'u S\'andor}
\def\corrAddress{
Institute for Theoretical Physics, Goethe University Frankfurt, 
	Max-von-Laue Str.~1, Frankfurt am Main, 60438, Germany}
\def\corrEmail{sandor@itp.uni-frankfurt.de}

\begin{document}
\onecolumn
\firstpage{1}

\title[The sensorimotor loop as a dynamical system]{The 
sensorimotor loop as a dynamical system: How regular motion 
primitives may emerge from self-organized limit cycles} 

\author[\firstAuthorLast ]{\Authors} 
\address{} 
\correspondance{} 

\extraAuth{}
\topic{Theory and Applications of Guided Self-Organization in Real and Synthetic Dynamical Systems}
\maketitle

\begin{abstract}
We investigate the sensorimotor loop of simple robots
simulated within the LPZRobots environment from the 
point of view of dynamical systems theory. For a
robot with a cylindrical shaped body and an actuator
controlled by a single proprioceptual neuron we find
various types of periodic motions in terms of stable
limit cycles. These are self-organized in the sense,
that the dynamics of the actuator kicks in only,
for a certain range of parameters, when the barrel 
is already rolling, stopping otherwise. The stability
of the resulting rolling motions terminates generally, 
as a function of the control parameters, at points
where fold bifurcations of limit cycles occur. We
find that several branches of motion types exist 
for the same parameters, in terms of the relative
frequencies of the barrel and of the actuator,
having each their respective basins of attractions
in terms of initial conditions. For low drivings
stable limit cycles describing periodic and drifting
back-and-forth motions are found additionally. These
modes allow to generate symmetry breaking explorative
behavior purely by the timing of an otherwise neutral 
signal with respect to the cyclic back-and-forth 
motion of the robot.
\tiny
 \keyFont{ \section{Keywords:} Sensorimotor loop, 
Adaptive behavior, Self-organization, Limit cycles, 
Period tripling, Embodiment, Explorative behavior, 
Symmetry breaking}
\end{abstract}

\section{Introduction}

Robots moving through an environment need to take the 
physical laws into account. This can be achieved
either via classical control theory \citep{de2012theory},
or by considering the full sensorimotor loop as an 
overarching dynamical system \citep{ay2012information}.
This distinction could be cast, alternatively, into
open-loop control, e.g.\ via central pattern generators 
\citep{ijspeert2008central}, and closed-loop schemes 
using feedback to control the states of an internal 
dynamical system \citep{dorf1998modern}. The presence 
of such feedback mechanisms capable of amplifying 
local instabilities are key components leading to the
emergence of self-organization \citep{der2012playful}.
A closely related notion is that of embodiment 
\citep{ziemke2003s}, for which no need arises for an 
explicit modeling of the interactions between the robot 
and its surroundings. The agent situated in a given environment
can be treated, in an embodied approach, as an overarching 
dynamical system, incorporating both the external dynamics 
(body-environment interaction), as well as the internal 
(controller-body) processes.
Thus, combining the closed-loop control with 
the embodied approach leads to movements generated through 
self-organizing processes. These, may in turn be guided 
by generic, e.g.\ information theoretical objective 
functions \citep{martius2013information}, such as predictive 
information \citep{ay2008predictive}, resulting in 
explorative or even playful behavior \citep{der2012playful}. 

Similar objective functions, such as the free energy 
\citep{friston2010free}, can also be considered for the 
brain as a whole \citep{baddeley2008information} and in 
the context of adaptive behavior \citep{friston2011free}.
Distinct control mechanisms for neural networks can also
be derived from other information theoretical generating 
functionals, such as the relative information entropy 
\citep{triesch2007synergies}, the mutual information 
\citep{toyoizumi2005generalized}, the Fisher information 
\citep{echeveste2014generating}, and the recently introduced 
active information storage measure
\citep{lizier2012local,dasgupta2013information}. 
Starting from first principles Hebbian learning rules 
have also been derived \citep{echeveste2015fisher}.

A parallel approach for studying the power of embodiment 
is provided by evolutionary robotics. Robots, selected 
through evolutionary processes \citep{nolfi2000evolutionary}
take environmental feedback naturally into account, as they
would otherwise not be positively selected. The notion of 
an acting agent in a reacting environment becomes blurry, 
to a certain extent, when the full sensorimotor loop is 
considered, with the motion coming to a standstill
without a fully functional feedback cycle. Within
other approaches to embodiment, the physical constraints 
acting on compliant real-world robots are studied
\citep{pfeifer2007self}, or the flow of information,
e.g.\ in terms of transfer entropy, through the
sensorimotor loop \citep{schmidt2013bootstrapping}.
A related question is how to ground actions generically,
i.e.\ without \'a priori knowledge, in sensorimotor 
perceptions \citep{olsson2006unknown}, or how to
select actions from universal and agent-centric measures 
of control \citep{klyubin2005empowerment}.

Abstracting from the sensorimotor loop, one may 
regard, from the point of view of dynamical system
theory \citep{beer2000dynamical}, motions as
organized sequences of movement primitives
in terms of attractor dynamics \citep{schaal2000nonlinear},
which the agent needs first to acquire by learning attractor 
landscapes \citep{ijspeert2002learning,ijspeert2013dynamical}.
These may be used later on for encoding the transients 
leading to periodic motions \citep{ernesti2012encoding}, 
or may furthermore self-organize into
complex behaviors \citep{tani2003self}. In this context
the fully embodied approach may serve as an algorithmic
first step to generate a palette of motion primitives.
One may also observe that all regular motions are, per 
definition, attractors in terms of stable limit cycles in 
the overarching sensorimotor loop, which may be controlled 
either actively \citep{laszlo1996limit}, or passively
in terms of limit-cycle walking \citep{hobbelen2008limit}. 
As an alternative approach for creating and controlling 
limit cycles one could use  prototype dynamical systems, 
a concept recently proposed for the study of complex 
bifurcation scenarios \citep{sandor2015general}.

\begin{figure}[t!]
\begin{center}
\includegraphics[height=0.3\textwidth]{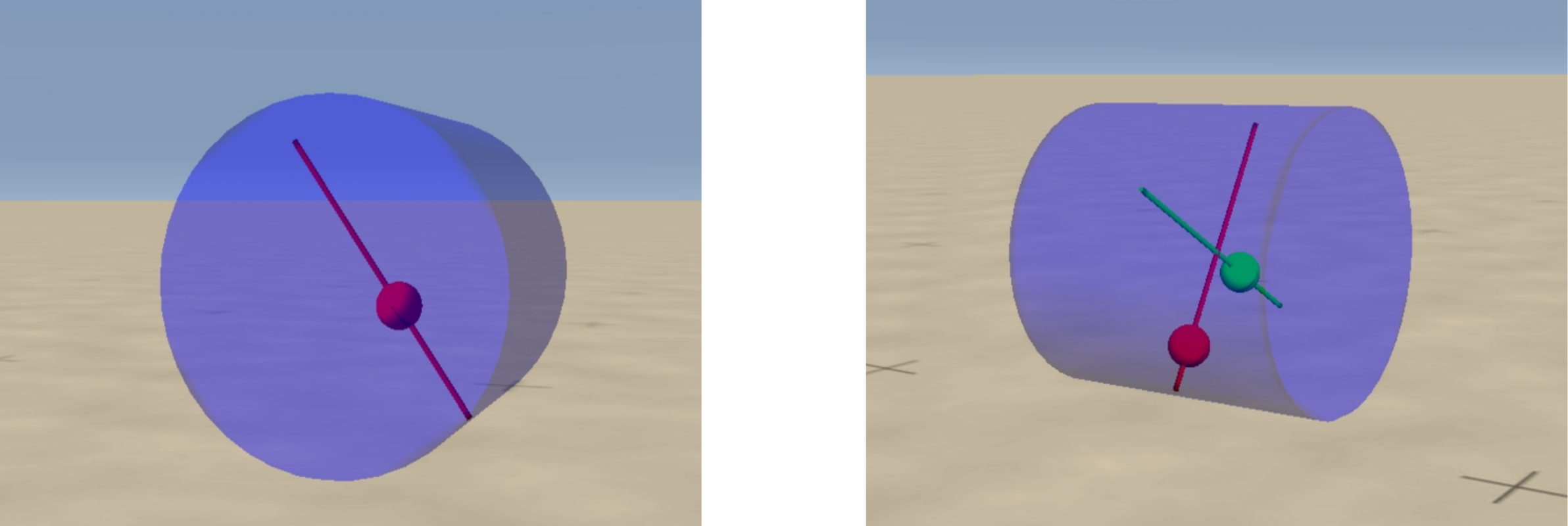}
\end{center}
\textbf{\refstepcounter{figure} 
\label{fig:barrels_lpz} 
Figure \arabic{figure}.}
{Screenshots from the LPZRobots simulation package, of the
one- and two rod barrel robots used (left- and right
panel respectively).}
\end{figure}

In the present study we examine in detail the notion 
of periodic movements as stable limit cycles,
using the LPZRobots package 
\citep{der2012playful,martius2013information}
for simulating robots (current development 
version), which are geometrically simple 
enough to allow for an at least partial modeling in 
terms of dynamical system theory \citep{gros2015complex}.
Our robots, see Figs.~\ref{fig:barrels_lpz} and 
\ref{fig:engine_barrel},
are controlled by a single proprioceptual
neuron with a time dependent threshold $b=b(t)$.
We find a region of parameters in which the motion
is fully embodied, and where the movement $v_b=v_b(t)$ of 
the robot and the threshold dynamics are mutually fully 
interdependent, vanishing when one of them, either
$b(t)$ or $v_b(t)$, is clamped. In engineering terms
the engine $db/dt$ powering the motion of the robot
is turned on dynamically through the feedback of its 
very motion.

We also find that a set of qualitatively distinct 
movements can arise for identical settings of the 
parameters in terms of stable limit cycles, having
their own distinct basins of attraction in phase space.
Control signals may hence switch between different motion 
primitives without the need to interfere with the
parameter setting of the sensorimotor loop. Most
modes found lead to regular motions with finite 
average velocities. We discovered however also a
particular mode corresponding to a cyclic 
back-and-forth movement, without an average 
translational motion of the robot. When the
parameter settings are changed in this mode, the
robot will enter a rolling motion, either to the
left or to the right, depending on the timing
of the signal with respect to the phase of the
cycle, allowing, as a matter of principle, for a 
truly explorative behavior.

A central result of the present study is that even
very simple controller dynamics (a single
differential equation, in our case) may lead via
the sensorimotor loop to surprisingly rich 
repertoires of regular motion primitives, which 
may be selected in turn through higher-order 
decision processes. This is due to the 
self-stabilization of motion patterns within the 
sensorimotor loop. Goal oriented behavior would
in this context be achieved not by optimizing motion
directly, but by selecting from the many attracting
states generated by an embodied controller within the 
overall sensorimotor loop.

\section{Methods}

We start by describing the one-neuron controller
used together with the actuator in terms of a damped 
spring, and the actual setup of the robot.

\subsection{Rate encoding neurons with internal adaption}
\label{sec:daption_rules}

In this paper we consider actuators controlled by
simple rate encoding neurons, characterized by  a 
sigmoidal transfer function
\begin{equation}
y(x,b) = \frac{1}{1+e^{a(b-x)}},
\qquad\quad
\dot b = \varepsilon a (2y-1)
\label{eq:sigmoidal}
\end{equation}
between the membrane potential $x$ and the firing
rate $y$, where $a$ is the gain, taken to be fixed,
and $b=b(t)$ a time-dependent threshold.
The dynamics $\dot b$ for the threshold in
(\ref{eq:sigmoidal}) would lead to $b\to x$
and $y\to 1/2$ for any constant input $x(t)=x$,
with a relaxation time being inversely proportional 
to the adaption rate $\varepsilon$. This adaption
rate can also be motivated by information-theoretical
considerations for the distribution of the firing
rates \citep{triesch2005gradient,markovic2010self}.

\begin{figure}[t!]
\begin{center}
\includegraphics[height=0.25\textwidth]{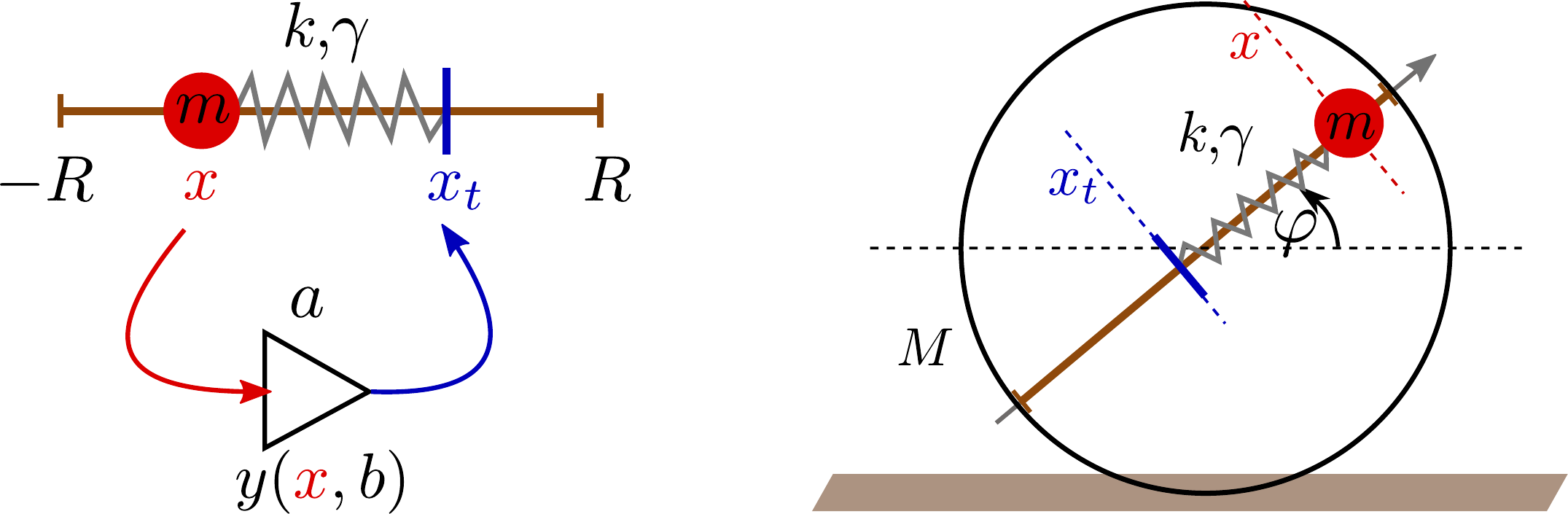}
\end{center}
\textbf{\refstepcounter{figure} 
\label{fig:engine_barrel} 
Figure \arabic{figure}.}
{\textit{Left:} Illustration of the proprioceptual single-neuron 
controlled damped spring actuator. The input $x$ of 
the neuron (described by Eq.~(\ref{eq:sigmoidal}))
is given by the actual position $x\in[-R,R]$ of the ball 
of mass $m$ moving on the rod, while the output $y$ being 
proportional, via Eq.~(\ref{eq:target_pos}), to the
target position $x_t$ of the ball. The PID controller
then simulates the dynamics of a damped spring, with constant
$k$ and damping $\gamma$, between the current and
the target positions of the mass.\newline
\textit{Right:} Sketch of the one-rod robot composed
of a barrel of mass $M$ and radius $R$, with a mass 
$m$ moving along a rod, as illustrated in the 
left panel. Slipping is not allowed, the robot  
moves hence with a velocity $v_b=R\omega=R\dot\varphi$, where 
$\varphi$ measures the angle of the rod with respect 
to the horizontal.}
\end{figure}

\begin{figure}[t]
\begin{center}
\includegraphics[height=0.45\textwidth]{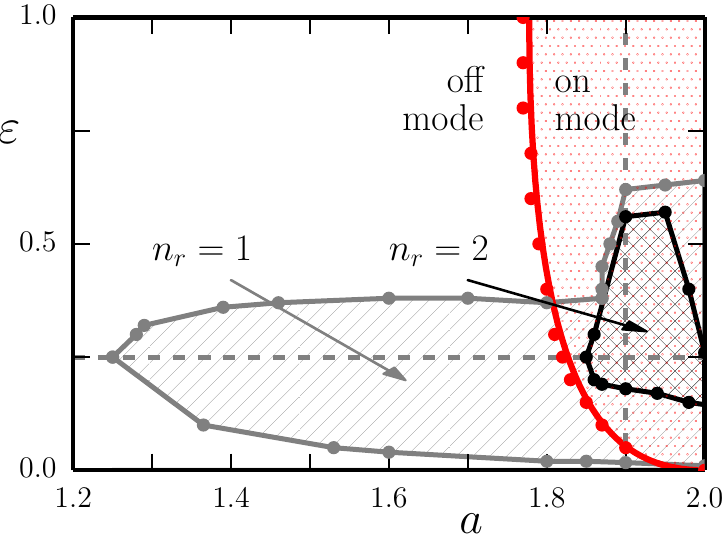} 
\end{center}
\textbf{\refstepcounter{figure} 
\label{fig:bif_horizontal} 
Figure \arabic{figure}.}
{The phase diagram of the one-rod barrel, as a function
of the gain $a$ and of the adaption rate $\varepsilon$. 
The results are obtained using the LPZRobots package, apart 
from the red solid line separating the off- and the on mode, 
which follows from (\ref{eq:horizontal}), for a fixed but 
otherwise arbitrary angle $\varphi=\varphi_0$. The dashed 
vertical and horizontal gray lines indicate the cuts used for
the phase diagrams presented in Fig.~\ref{fig:one_axis_eps_a}.
The number of stable limit cycles found in the respective 
parameter regions are denoted by $n_r$. 
\newline
\textit{Non-rolling modes:} The red dots/line 
indicate the locus of a Hopf bifurcation, where a stable non-rolling 
limit cycle (on mode) emerges from the trivial non-rolling fixpoint 
(off mode). In the off mode the 'engine' $db(t)/dt$, see
Eq.~(\ref{eq:sigmoidal}), kicks in only when the barrel 
is already moving. 
\newline
\textit{Rolling modes:} 
Shown are the regions containing $n_r=1$ (enclosed by the solid gray 
line) and $n_r=2$ (enclosed by the solid black line)  
attracting limit-cycles corresponding to a barrel moving with 
a finite velocity $\langle v_b\rangle$. Note, that the robot
is able to move also in the off mode (of the engine).
The stationary and the drifting back-and-forth modes, discussed
in Fig.~\ref{fig:one_axis_forth_back_modes}, have been
omitted, in order to avoid overcrowding.
}
\end{figure}

\subsection{Damped spring actuators}
\label{sec:engine}

Our robots are controlled by actuators regulating
the motion of the ball of mass $m$ on a rod, as 
illustrated in Fig.~\ref{fig:engine_barrel}, from
its actual position $x$ on the rod, to its target
position 
\begin{equation}
\begin{aligned}
x_t  & = 2R\left(y(x,b)-\frac{1}{2}\right)\,,
\end{aligned}
\label{eq:target_pos}
\end{equation}
where $R$ is the radius of the barrel containing the
rod and where $y(x,b)$ is the sigmoidal (\ref{eq:sigmoidal}).
We note that the input and the output of the neuron are, via
(\ref{eq:target_pos}), of the same dimensionality, namely
positions. The force $F=m\ddot x$ moving the ball is
evaluated by the PID controller
\begin{equation}
F = g_P(x_t-x) + g_I\int_0^t(x_t-x)dt + g_D\frac{d(x_t-x)}{dt}\,,
\label{eq_PID}
\end{equation}
provided by the LPZRobots simulation environment
\citep{der2012playful}, characterized by the standard 
PID-control parameters $g_P$, $g_I$ and $g_D$. 

For our simulations we considered the case $g_I=0$, for which
the PID controller reduces to a damped spring, see
Fig.~\ref{fig:engine_barrel},
\begin{equation}
\begin{aligned}
m\ddot{x}  & = -k(x-x_t)-\gamma\frac{d(x-x_t)}{dt},
\end{aligned}
\label{eq:damped_oscillator}
\end{equation}
with $k=g_P$ and $\gamma=g_D$.
\begin{itemize}
\item Eq.~(\ref{eq:damped_oscillator}) represents only the
  contribution of the actuator to the force moving the ball
  along the rod. The gravitational pull acting on the mass 
  $m$, and the centrifugal force resulting from the rolling 
  motion of the barrel on the ground are to be added to the 
  RHS of Eq.~(\ref{eq:damped_oscillator}).
\item The target position $x_t=x_t(t)$ is time-dependent
  through (\ref{eq:target_pos}) and (\ref{eq:sigmoidal}).
\item Eq.~(\ref{eq:damped_oscillator}) is strictly dissipative,
  due to the damping $\gamma>0$. The same holds for the rolling
  motion of the barrel on the ground, which is also characterized
  by a finite rolling friction. Thus, the dynamics $d b/d t$ of 
  the threshold in (\ref{eq:sigmoidal}) can be considered as an engine,
  providing, by adjusting continuously the target position $x_t$ 
  of the ball, and hence the length of the spring, the energy 
  dissipated by the physical motions.
\end{itemize}

\begin{figure}[t!]
\begin{center}
\includegraphics[width=0.95\textwidth]{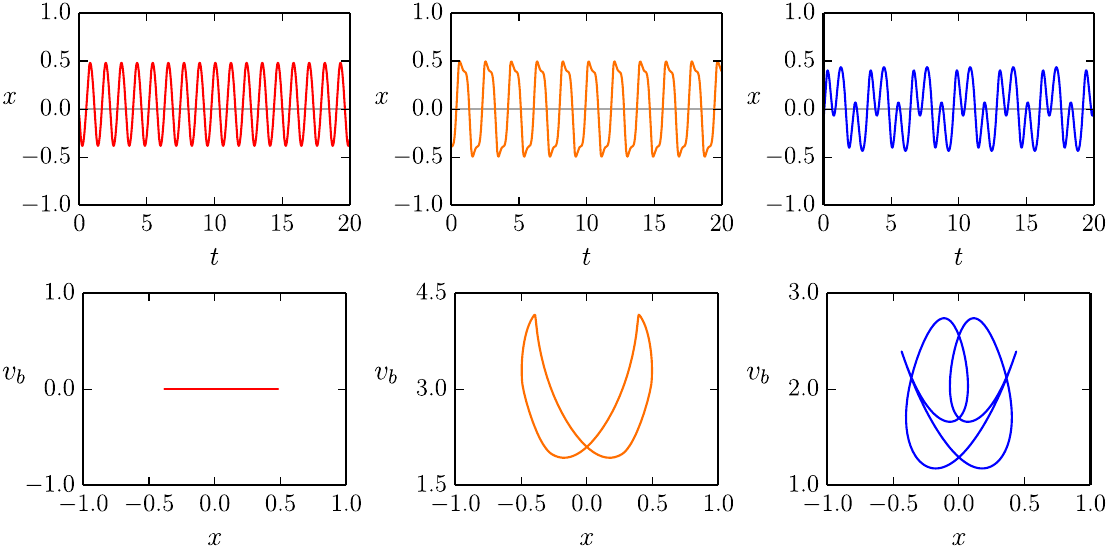}
\end{center}
\textbf{\refstepcounter{figure} 
\label{fig:one_axis_modes} 
Figure \arabic{figure}.}
{The motion $x(t)$ of the mass along the rod of the one-rod
barrel (top row), together with the corresponding phase-plane 
trajectories $(x(t),v_b(t))$ (bottom row), compare 
Fig.~\ref{fig:engine_barrel}. The gain and the adaption rate 
are $a=1.9$ and $\varepsilon=0.25$ respectively. Shown are 
the 0:1, 1:1 and 1:3 modes (left/middle/right column).
Note, that the velocity $v_b(t)$ of the barrel vanishes
for the 0:1 mode, oscillating but remaining otherwise
positive for the 1:1 and the 1:3 mode.
(\href{https://www.youtube.com/watch?v=bF7298z_iv8&t=29s}
      {click for movie}).
}
\end{figure}
 
\subsection{Motion of a Mass on a Fixed Rod}
\label{sec:fixedRod}

As an example we consider a robot, for which we keep
the angle $\varphi$ between the rod and the horizontal
fixed, $\varphi=\varphi_0$, by preventing it from rolling. 
We are then left with a self-coupled motion of a ball 
along a rod, as illustrated in the left panel of 
Fig.~\ref{fig:engine_barrel}, resulting in a dynamics
similar to the one of a self coupled neuron 
\citep{markovic2012intrinsic,gros2014attractor}.
Using $\Omega^2={k}/{m}$ and $\Gamma={\gamma}/{m}$
we find in this case 
\begin{equation}
\begin{array}{rclcrcl}
\dot{x} &=& v 
&\qquad\quad&
\dot{x}_t& =& 2\,R\,a\,y(1-y)(v-\dot{b})\\
\dot{v} &=& -\Omega^2(x-x_t) - \Gamma (v-\dot{x}_t)
            -g\sin(\varphi_0)
&\qquad\quad&
\dot{b} &=& 2\,\varepsilon\,a (y-1/2)
\end{array}~,
\label{eq:horizontal}
\end{equation}
when combining Eqs.~(\ref{eq:sigmoidal}), (\ref{eq:target_pos})
and (\ref{eq:damped_oscillator}). The gravitational term $-g\sin(\varphi_0)$ 
can be transformed away via
\begin{equation}
x \rightarrow x-{g}/{\Omega^2}\sin\varphi_o\,, \qquad
b \rightarrow b-{g}/{\Omega^2}\sin\varphi_o\,,
\end{equation}
and does hence not influence the phase diagram, 
which is shown in Fig.~\ref{fig:bif_horizontal}
for $\Omega^2=200$, $\Gamma=2\Omega$ and $g=9.81$.
We have used standard numerical methods \citep{clewley2012hybrid}.

We find a Hopf bifurcation line separating the stability 
regions for the trivial fixpoint and for a limit cycle, 
denoted respectively as off and on modes. 
This behavior is similar to the
one observed for a self coupled neuron with
intrinsic adaption \citep{markovic2012intrinsic,gros2014attractor}.

\section{Results}

In Fig.~\ref{fig:barrels_lpz} the screenshots of the one- and 
two-rod robots simulated with the LPZRobots package 
(current development version)
\citep{der2012playful,martius2013information} are presented.
Throughout
the simulations the control parameters $\Gamma=2\Omega$ 
and $\Omega^2=200$ for the actuator, $\Lambda=1$ for the  
mass ratio $m/M$ (ball to barrel), $R=1$ for the radius for 
the barrel, and $\Psi=0.3$ for the coefficient of the 
rolling friction have been held constant, varying only 
the adaption rate $\epsilon$ for the threshold of 
the neuron, and the gain $a$. For the simulations a 
stepsize of $0.001$ was used. In the Figures 
(and in the rest of the paper) the 
parameters will be presented in dimensionless units, with 
SI units being implied: seconds/meter for the 
time and length respectively and $g=9.81\,\mathrm{m}/\mathrm{s}^2$ 
for the gravitational acceleration. Our barrel has a radius of 
$1\,\mathrm{m}$ and a moving mass of $1\,\mathrm{kg}$, 
rolling typically at speeds of 
$(1-4)\,\mathrm{m}/\mathrm{s}\approx (3-12)\,\mathrm{km}/\mathrm{h}$.
A table of the parameters is given in the Supplementary Material.

\subsection{One-rod barrel}

The overall phase diagram of the one-rod barrel
shown in Fig.~\ref{fig:bif_horizontal} contains
regions of non-rolling fixpoints or limit cycles, 
and regions where one or more
attracting limit cycles corresponding to a continuously 
rolling barrel are present, in part additionally. Depending
on the initial conditions the system will eventually
settle into one of the attracting states. 

\begin{figure}[t!]
\begin{center}
\includegraphics[height=0.35\textwidth]{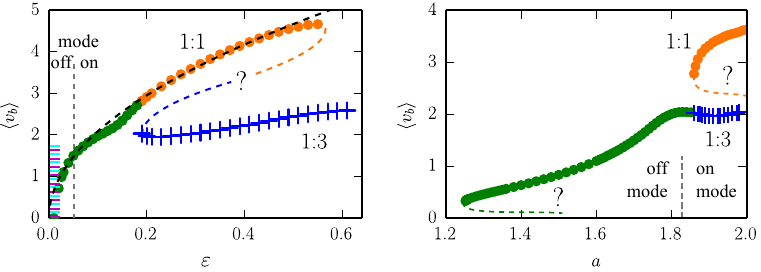} 
\end{center}
\textbf{\refstepcounter{figure} 
\label{fig:one_axis_eps_a} 
Figure \arabic{figure}.}
{The average speed $\langle v_b\rangle$ of the one-rod barrel 
for the 1:1 (green/orange dots) and for 1:3 (blue crosses) 
mode. The vertical dashed line denotes the locus of the 
Hopf bifurcation line shown in Fig.~\ref{fig:bif_horizontal}. 
In the off mode (on mode) the attracting state for the
non-rolling mode is a stable fixpoint (limit cycle)
respectively. Presumably existing unstable limit cycles 
are indicated by dashed lines (labeled with question 
marks).\newline
\textit{Left:} For a gain $a=1.9$. The colored region for 
very small adaption rates $\varepsilon$ indicates 
a region with both stable and drifting back-and-forth 
modes, further described in 
Fig.~\ref{fig:one_axis_forth_back_modes}. \newline
\textit{Right:} For an adaption rate $\varepsilon=0.25$.
}
\end{figure}

\subsubsection{Coexisting modes as behavioral primitives.} 

Standard robot control aims at achieving a predefined 
outcome and for this purpose it is indispensable, that
identical robot actions lead also to identical movements. 
This is not necessarily the case for robots controlled
by self-organized processes, as investigated here.

In Fig.~\ref{fig:one_axis_modes} we illustrate the
time series and the corresponding phase-space plots
of the dominant modes of the one-rod barrel shown in
Fig.~\ref{fig:barrels_lpz}. The simulation parameters
$a=1.9$ for the gain, and the $\varepsilon=0.25$ adaption 
rate are close to the Hopf bifurcation line shown in 
Fig.~\ref{fig:bif_horizontal}, but in the on mode.
Which means, that the ball moves both for fixed
horizontal or vertical rods.

The first of the three coexisting stable limit cycles, 
illustrated in Fig.~\ref{fig:one_axis_modes}, corresponds 
to the non-moving barrel with the ball oscillating vertically
along the rod (first column). For the second, 1:1 mode, the
average rolling frequency of the barrel and of the
oscillation of the ball along the rod match (second column).
For the 1:3 mode the corresponding ratio of frequencies
is however 1:3 (third column).

The occurrence of several distinct limit cycles for
identical parameters can be interpreted in terms of
behavioral primitives, potentially allowing an agent 
to switch rapidly between different types of motions,
by shortly destabilizing the currently active limit
cycle.

Note that the self-coupled neuron, controlling 
the dynamics of the ball along the horizontally fixed rod, 
has only two possible stable states (a fixpoint and a limit 
cycle). Considering however the fully embodied rolling 
robot, coexisting states are arising, which can 
lead to different behavioral patterns purely as a result 
of the environmental context. An external force applied 
to the robot can qualitatively change its behavior, 
indicating the sign of multifunctionality 
\citep{williams2013environmental}.

\subsubsection{Embodiment as self-organized motion.} 

Most robots are autonomously active in the sense, that
the motion is not essentially dependent on the feedback
of the environment. For the case of self-organized motion,
as considered here, there would be, on the other side,
no motion when the sensorimotor loop would be interrupted.

We present in the left plot of Fig.~\ref{fig:one_axis_eps_a}
the evolution of the self-sustained rolling modes, in terms
of the averaged measured velocity, for $a=1.9$ and as a 
function of adaption rate $\varepsilon$. The dashed black 
line indicates, as a guide to the eye, that the velocity 
increases roughly $\propto\sqrt{\varepsilon}$ for the
1:1 mode. The two branches are stable
for $\varepsilon\in[0.018,0.55]$ and 
$\varepsilon\in[0.19,0.61]$ respectively for the
1:1 and the 1:3 mode, and terminate (presumably)
through saddle node bifurcations of limit cycles.
We have indicated this scenario by adding by
hand in Fig.~\ref{fig:one_axis_eps_a}, as
guides to the eye, the respective unstable
branches.

The locus of the Hopf bifurcation shown in 
Fig.~\ref{fig:bif_horizontal}, at $\epsilon\approx0.05$, 
is indicated in (the left panel of) 
Fig.~\ref{fig:one_axis_eps_a} by the dashed vertical line, 
separating the off from the on mode. In the off and 
on modes the non-rolling attractors are a fixpoint and 
a limit cycle respectively. Note that self-sustained 
rolling modes exist in the off mode 
as well, where the 'engine' $db(t)/dt$ of the barrel 
only kicks in, through amplifying local fluctuations 
(damped oscillations around the fixpoint), 
when the barrel is already moving. 
This underlines the embodied nature of the motion, 
which arises in a truly self-organized fashion (in
term of dynamical systems theory \citep{gros2015complex})
through the bidirectional feedback between environment
and both the body and the controller of the robot.

\begin{figure}[t!]
\begin{center}
\includegraphics[width=0.95\textwidth]{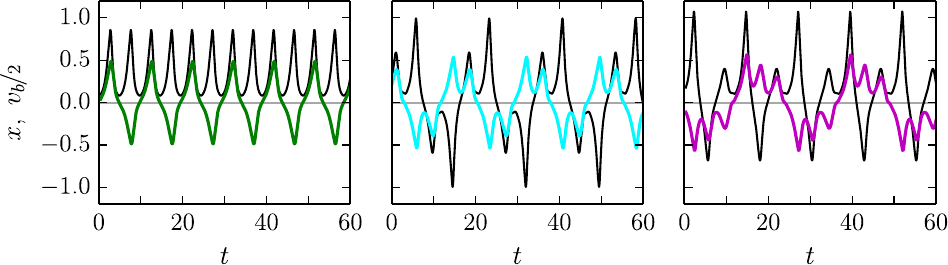}
\end{center}
\textbf{\refstepcounter{figure} 
\label{fig:one_axis_forth_back_modes} 
Figure \arabic{figure}.}
{The time evolution of the  position $x$ (colored lines) 
of the mass along the rod, and of the (rescaled) speed $v_b$ 
of the barrel (black lines), for $a=1.9$ and 
$\varepsilon=0.019/0.017/0.015$ 
(\textit{left/center/right}), all in the off mode
(compare Fig.~\ref{fig:one_axis_eps_a}).
The respective average 
velocities are $\langle v_b\rangle=0.63/0.00/0.25$
for the 1:1 mode (left), the stationary back-and-forth
mode (middle) and the drifting back-and-forth mode (right).
(\href{https://www.youtube.com/watch?v=7KxlddLiVa4&t=34s}
      {click for movie}).
}
\end{figure}

However, in the absence of feedback mechanisms (such as 
centrifugal- and Coriolis-forces), the neuron controlled 
actuator could only generate a single regular rolling 
motion, similar to the ones achieved by sending motor 
signals generated by some central pattern generators 
\citep{der2012playful}. This is not the case for our
robot, which exhibits, as shown in 
Fig.~\ref{fig:one_axis_eps_a} (and in
Fig.~\ref{fig:one_axis_forth_back_modes}, see discussion
below) a wide spectrum of possible rolling modes.

\subsubsection{Avoided pitchfork bifurcations of limit cycles} 

In the right panel of Fig.~\ref{fig:one_axis_eps_a} we present
the measured mean velocity $\langle v_b\rangle$ of the ball
for $\varepsilon=0.25$, as a function of the gain $a$. 
The Hopf bifurcation between the off- and on- 
non-rolling modes occurs at $a_H\approx1.83$, compare 
Fig.~\ref{fig:bif_horizontal}.

For $1.23<a<1.83$ the ball hence is moving in the off mode,
with the engine kicking in only through the feedback from
the environment, which we interpret as self-organized
embodied motion, with the environment being an essential 
component of the overarching dynamical system. 

Comparing both panels of Fig.~\ref{fig:one_axis_eps_a}
one can notice that the low-velocity mode (green dots) 
connects either to the the 1:1 mode (as in the left panel)
or to the 1:3 mode (as in the right panel). The reason 
for the apparent discrepancy lies in the fact, that 
the respective bifurcation line is oblique in the 
phase space plane $(a,\varepsilon)$. The evolution of 
these modes suggests in any case, that the low-velocity 
mode connects to the two higher-velocity modes via an avoided
pitchfork transition of limit cycles \citep{gros2015complex}.

\subsubsection{Explorative motion via noise induced 
directional switching.} 

Our robot contains a single dynamical variable, the
threshold $b(t)$, generating self-stabilizing motions 
via the sensorimotor loop. The palette of modes generated
is, despite this apparent simplicity, surprisingly large
and may be used to generate higher order behavior.

There are three dominant branches, the 0:1, 1:1 and 1:3 
modes (in terms of the ratios of the respective barrel- and 
mass frequencies), compare Figs.~\ref{fig:one_axis_modes}
and \ref{fig:one_axis_eps_a}, which are stable for a wide 
range of parameters. We found in addition also a parameter 
region for which different types of motions arise 
from minute changes of control parameters, such as 
the adaption rate $\varepsilon$. 

In Fig.~\ref{fig:one_axis_forth_back_modes}
the motion $x(t)$ of the ball along the rod and
the velocity $v_b(t)$ of the barrel are given
for three closely spaced adaption rates
$\varepsilon=0.019,0.017$ and $0.015$,
for which three qualitatively different types
of motions are found (which have partially,
but not completely overlapping stability regions).
\begin{itemize}
\item For $\varepsilon=0.019$ the standard 1:1 
rolling motion is recovered, with an average
velocity $\langle v_b\rangle=0.63$.
\item For $\varepsilon=0.017$ a new mode arises,
for which the ball rolls back and forth forever.
The motion is exactly symmetric with respect to the
left and to the right, and the average
velocity $\langle v_b\rangle=0.0$ of the barrel 
hence vanishes exactly.
\item For $\varepsilon=0.015$ the ball also rolls
back and forth, but asymmetrically, giving rise
to a drifting motion with small but finite average
velocity of $\langle v_b\rangle=0.25$.
\end{itemize}

\begin{figure}[t!]
\begin{center}
\includegraphics[width=0.85\textwidth]{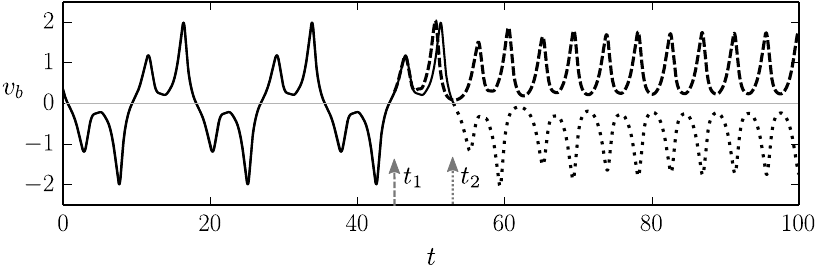}
\end{center}
\textbf{\refstepcounter{figure} 
\label{fig:change_direction} 
Figure \arabic{figure}.}
{Two superimposed runs for the time evolution of the 
speed $v_b$ of the barrel (black lines), for $a=1.9$. 
In the first run the adaption rate $\varepsilon$ is 
changed discontinuously at time $t_1=45$ from 
$\varepsilon=0.017$ (corresponding to the stationary 
back-and-forth mode, see 
Fig.~\ref{fig:one_axis_forth_back_modes})
to $\varepsilon=0.02$ (corresponding to the
1:1 rolling mode). In the second run identical
initial conditions have been used and an identical
change is made to the adaption rated $\varepsilon$,
but now at time $t_2=53$. In both runs (dashed and dotted
lines respectively), the barrel settles into the 1:1
rolling motion, albeit in opposite directions
(to the right/left with $\langle v_b\rangle>0$
and $\langle v_b\rangle<0$ respectively).
(\href{https://www.youtube.com/watch?v=L9dS1Kw_TtI&t=11s}
      {click for movie}).
}
\end{figure}

The occurrence of a limit cycle corresponding to a 
symmetric back-and-forth rolling motion, sandwiched
between symmetry breaking modes, gives rise to an 
interesting venue for the generation of explorative 
behaviors, as the robot will be sensitive to finite
but otherwise very small perturbations influencing 
its internal control parameters. This behavior
is illustrated in Fig.~\ref{fig:change_direction}.
Depending on the
timing of the perturbation with respect to the 
back-and-forth rolling cycle, the robot will settle 
into a left- or into a right-moving motion (in the 1:1 or 
in the back-and-forth drifting mode respectively for
increasing/decreasing $\varepsilon$).
It is hence possible to break spatial symmetries,
in general, purely via the timing of a perturbation.
The perturbation itself, here acting on the adaption 
rate $\varepsilon$, does not need to carry any
information about the direction of motion.

\subsection{Two-rod barrel}

Adding a second actuator perpendicular to the first 
one, a neuron controlled ball moving along a rod, one 
can increase the complexity of the robot (see the right 
picture of Fig.~\ref{fig:barrels_lpz}). Both actuators 
work, in our setup, independently, with the crosstalk
being provided exclusively by the environmental feedback. 
Both actuators are identical to the rod used for the 
single-rod barrel, with each rod having its own adapting 
threshold $b_\alpha(t)$ and membrane potential 
$x_\alpha(t)$, with $\alpha=1,2$. The adaption rate
$\varepsilon$, the gain $a$, and all other parameters
are identical for the two rods.

\begin{figure}[t]
\begin{center}
\includegraphics[height=0.35\textwidth]{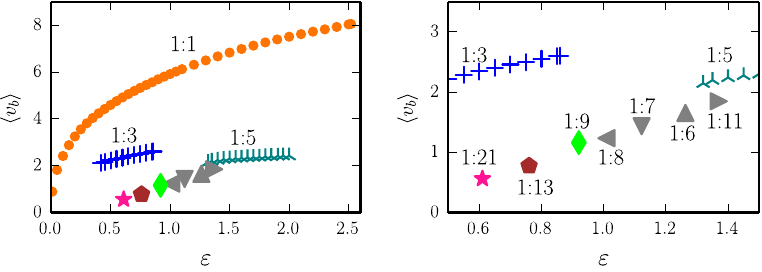}
\end{center}
\textbf{\refstepcounter{figure}
\label{fig:twoaxes_barrel_eps_v} 
Figure \arabic{figure}.}
{\textit{Left:} The average speed $\langle v_b\rangle$ 
of the two-rod barrel for the 1:1/1:3/1:5 
(orange dots, blue crosses, dark-cyan stars) modes. 
The gain is $a=1.9$, all other parameters are identical
to the ones used for the one-rod barrel. The respective
time series and phase-space plots are presented in 
Fig.~\ref{fig:twoaxes_barrel_time_series}. The 
filled symbols denote examples of additional higher 
order modes, of which the 1:21,1:13 and 1:9 
(pink star, maroon pentagon, green rhombus) 
are illustrated in Fig.~\ref{fig:two_axis_higher_modes}.
\textit{Right:} A blow up, showing the relative location 
of the 1:8,1:7,1:6 and 1:11 modes found at 
$\varepsilon$=1.009,1.122,1.263 and 1.370 respectively.
}
\end{figure}

In Fig.~\ref{fig:twoaxes_barrel_eps_v} we show in 
the right panel the stability range, for $a = 1.9$ 
and as a function of the adaption rate $\varepsilon$, 
of the three most dominant rolling modes 
(1:1, 1:3 and 1:5) of the two-rod barrel. 
A large variety of higher order 1:M modes 
(with M being an integer) is found in addition. 
We did not carry out a systematic search of their 
stability range, which becomes progressively smaller 
with increasing M, and present here only exemplary 
parameter settings for which the respective modes 
have been found by trial-and-error (by randomly kicking 
the barrel). A blow-up is given in the right panel of 
Fig.~\ref{fig:twoaxes_barrel_eps_v}. Most values of M
found are odd, but not exclusively. We cannot exclude, 
at this stage, that an infinite cascade $M\rightarrow\infty$  
of higher order limit cycles may possibly occur.

\begin{figure}[t!]
\begin{center}
\includegraphics[width=0.95\textwidth]{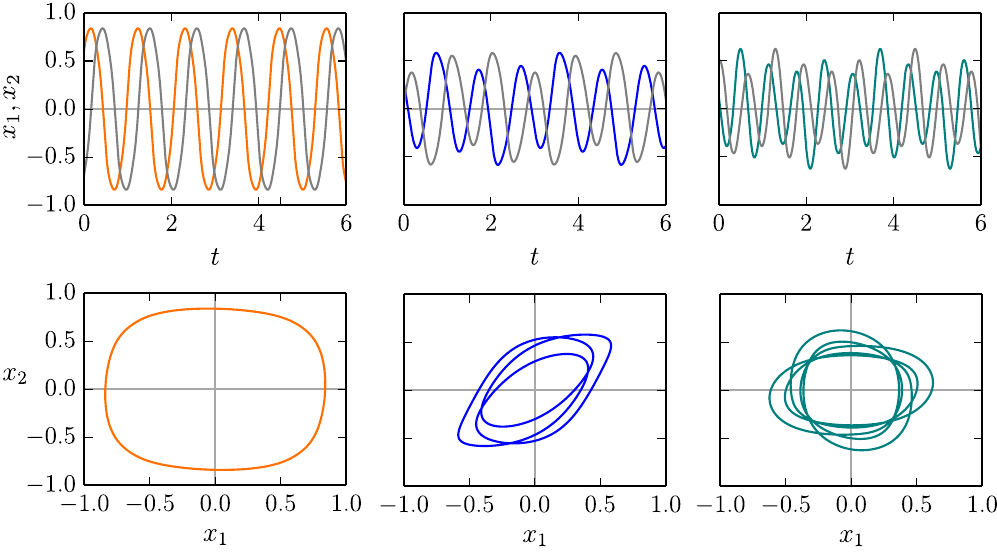}
\end{center}
\textbf{\refstepcounter{figure}
\label{fig:twoaxes_barrel_time_series} 
Figure \arabic{figure}.}
{Time series $x_1(t)$ and $x_2(t)$ of the balls along
the two rods of the two-rod barrel (top row), and the
respective phase plots $(x_1(t),x_2(t))$. Shown are the
1:1/1:3/1:5 modes (left/middle/right column) for 
$\varepsilon=1.0/0.5/1.5$, compare
Fig.~\ref{fig:twoaxes_barrel_eps_v}, needing
respectively 1/3/5 revolutions around the origin
$(x_1,x_2)=(0,0)$ in order to close.
(\href{https://www.youtube.com/watch?v=HIGPPKOCxQU&t=14s}
      {click for movie}).
}
\end{figure}

\begin{figure}[t!]
\begin{center}
\includegraphics[width=0.95\textwidth]{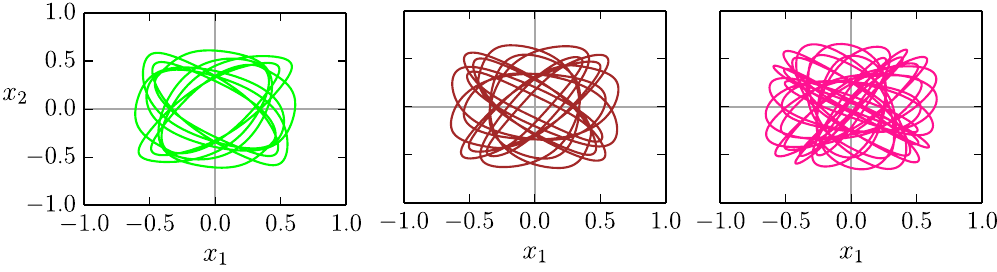}
\end{center}
\textbf{\refstepcounter{figure}
\label{fig:two_axis_higher_modes}
Figure \arabic{figure}.}
{Examples of higher order limit cycles found for the 
two-rod barrel, closing (within numerical accuracy,
viz the thickness of the lines)
after 9/13/21 revolutions around the origin 
$(x_1,x_2)=(0,0)$ (left/middle/right). The gain is 
$a=1.9$ and the respective adaption rates are
$\varepsilon=0.61,0.76$ and $0.92$, compare 
Fig.~\ref{fig:twoaxes_barrel_eps_v}.}
\end{figure}

The time series and the respective phase space trajectories 
$(x_1(t), x_2(t))$ of the 1:1, 1:3 and of the 1:5
modes are presented in Fig.\ref{fig:twoaxes_barrel_time_series}. 
As one can see in the time series plots, the two independent 
actuators, being only coupled through the dynamics of the barrel, 
self-organize themselves in a constant phase-shift, 
necessary for a consistent rolling.
In the reduced phase space $(x_1 , x_2)$ the trajectories 
exactly close on themselves, needing respectively 1,3 and 5 
revolutions around the origin $(0,0)$ to close, for 
respectively the 1:1, 1:3 and for the 1:5 limit cycles. 
In Fig.~\ref{fig:two_axis_higher_modes} we show the 
corresponding phase-space trajectories of the M=9, 13
and 21 limit cycles. These modes have progressively slower 
average velocities $v_b$ , compare Fig.~\ref{fig:twoaxes_barrel_eps_v}, 
and smaller basins of attractions, being otherwise 
regular stable limit cycles. Whether they arise through a
bifurcation cascade of limit cycles 
\citep{sandor2015general}, or via some other mechanism, 
is however beyond the scope of the present study.

\section{Discussion}

It is, in a certain sense, a trivial statement, that
the environment is part of the dynamical system a
biological or artificial agent lives in. Little of
the environmental dynamics is however in general
accessible, or known, from the perspective of a 
robot, and it is hence often more suitable, as
in closed-loop control \citep{dorf1998modern},
to consider the sensorimotor loop as a sequence of 
stimulus-response reactions of the agent, eliciting at 
every step the subsequent environmental signal.
Here we have considered 
simple barrel-shaped robots in a simulated environment, 
for which the sensorimotor loop constitutes truly a
dynamical system, capable of generating, even in a 
simple setup, a very rich palette of dynamical modes and 
hence a wide range of qualitatively different types of 
motions.

The dominant rolling modes found are 1:M attractors, 
where the actuators cycle M=1,3,5,..  times during 
one revolution $\varphi\to\varphi+2\pi$ of the barrel. 
These modes coexist with non-rolling modes, having 
their own respective basins of attractions, emerging 
from the mutual feedback of robot and environment. There 
exist, in addition, regions of phase space with
stationary rolling modes (rolling periodically
back and forth), and drifting back-and-forth modes. 
We have also found preliminary indications of rolling 
modes living on two- or higher dimensional tori, with 
incommensurate revolution frequencies, which we did
however not investigate in detail in the present study. 
There may additionally exist further attracting states, 
yet not discovered when performing numerical simulations 
within the LPZRobots environment.

All modes found are attracting dynamical states and
hence robust against noise. This robustness varies 
however, with the dominant 1:1 being the most stable,
and higher order modes, like the 1:3 or the 1:21 
limit cycles, being relatively less stable. There is, 
in addition, the need to overcome the dissipation, 
which is present in the simulated environment, by an
appropriate energy intake of the actuator. As for
all robots the question then arises, whether the
observed behavior can be considered as dominantly 
driven, in the sense of actuator overpowering,
or as self-organized, via an inherent and essential
feedback loop through the environment (in this context 
see \citep{egbert2010minimal} for an analogous discussion
in the context of bacterial sensorimotor system involving
chemotaxis).

Actuator-controlled behavior would generally lead,
in our perspective, to rather stereotypical movements
modes. The fact that our robots show a very large
variety of modes upon changing the adaption
rate $\varepsilon$, viz the reaction time $1/\varepsilon$
of the actuator, indicates self-organization. These
modes are also partially overlapping with several
rolling modes possibly coexisting for the same settings. 
It is then a question of starting conditions, into which
behavior the robot then settles.

We have also investigated the dynamics of the actuators
employed, a damped-spring ball moving along a rod, when 
the rolling motion $d\varphi/dt\to0$ of the barrel is turned 
off. In this setting the environmental feedback from the rolling
motion is not present. We find parameter regions where the 
engine is autonomously active and parameter regions, where
the engine shuts itself off. In the later region the engine 
may be kicked in again, when the barrel is given a kick, and 
allowed to roll normally. In this case the environmental
feedback is hence essential, and the motion of the robot
is a consequence of self-organizing processes in the 
combined phase space of the internal degrees of freedom
of the robot and of the physical environment.

Thus the behavior of the robot can not be attributed to merely
one of the subsystems, but it is a property of the 
coupled brain-body-environment system, a result also found 
in the context of minimally cognitive agents 
\citep{beer2003dynamics,beer2015information}.
Since we are not aiming here for the presence of higher 
level cognitive processes, our work can be seen as a 
purely dynamical systems approach for understanding 
embodiment directly within the sensorimotor loop.

Our work has been performed with the LPZRobots simulation
package, which has been used extensively to investigate 
the emergence of `playful' behavior and sensorimotor 
intelligence in terms of intermittent chaotic motion 
patterns \citep{der2012playful,martius2013information}. 
In this context our investigation is embedded in the
long-standing effort 
\citep{taga1991self,kelso1994informational,pfeifer2007self,der2015novel} 
to reduce the demanding problem of programming robots by 
investigating the emergence of self-organized motions 
within the sensorimotor loop.

\section*{Conflict-of-Interest Statement}
The authors declare that the research was conducted 
in the absence of any commercial or financial 
relationships that could be construed 
as a potential conflict of interest.

\section*{Author Contributions}

Most data and figures where produced by B.~S\'andor,
the paper written by C.~Gros and B.~S\'andor, with
T.~Jahn and L.~Martin adding data and material.

\section*{Acknowledgments}

We thank Georg Martius for extensive discussions 
and for helping setting-up the LPZRobots simulation
environment.


\bibliographystyle{frontiersinSCNS&ENG}
\bibliography{sensorimotor_loop}

\begin{thebibliography}{44}
\providecommand{\natexlab}[1]{#1}
\expandafter\ifx\csname urlstyle\endcsname\relax
  \providecommand{\doi}[1]{doi:\discretionary{}{}{}#1}\else
  \providecommand{\doi}{doi:\discretionary{}{}{}\begingroup
  \urlstyle{rm}\Url}\fi
\providecommand{\selectlanguage}[1]{\relax}
\providecommand{\bibAnnoteFile}[1]{%
  \IfFileExists{#1}{\begin{quotation}\noindent\textsc{Key:} #1\\
  \textsc{Annotation:}\ \input{#1}\end{quotation}}{}}
\providecommand{\bibAnnote}[2]{%
  \begin{quotation}\noindent\textsc{Key:} #1\\
  \textsc{Annotation:}\ #2\end{quotation}}

\bibitem[{Ay et~al.(2012)Ay, Bernigau, Der, and Prokopenko}]{ay2012information}
Ay, N., Bernigau, H., Der, R., and Prokopenko, M. (2012).
\newblock Information-driven self-organization: the dynamical system approach
  to autonomous robot behavior.
\newblock \emph{Theory in Biosciences} 131, 161--179
\bibAnnoteFile{ay2012information}

\bibitem[{Ay et~al.(2008)Ay, Bertschinger, Der, G{\"u}ttler, and
  Olbrich}]{ay2008predictive}
Ay, N., Bertschinger, N., Der, R., G{\"u}ttler, F., and Olbrich, E. (2008).
\newblock Predictive information and explorative behavior of autonomous robots.
\newblock \emph{The European Physical Journal B} 63, 329--339
\bibAnnoteFile{ay2008predictive}

\bibitem[{Baddeley et~al.(2008)Baddeley, Hancock, and
  F{\"o}ldi{\'a}k}]{baddeley2008information}
Baddeley, R., Hancock, P., and F{\"o}ldi{\'a}k, P. (2008).
\newblock \emph{Information theory and the brain} (Cambridge University Press)
\bibAnnoteFile{baddeley2008information}

\bibitem[{Beer(2000)}]{beer2000dynamical}
Beer, R.~D. (2000).
\newblock Dynamical approaches to cognitive science.
\newblock \emph{Trends in cognitive sciences} 4, 91--99
\bibAnnoteFile{beer2000dynamical}

\bibitem[{Beer(2003)}]{beer2003dynamics}
Beer, R.~D. (2003).
\newblock The dynamics of active categorical perception in an evolved model
  agent.
\newblock \emph{Adaptive Behavior} 11, 209--243
\bibAnnoteFile{beer2003dynamics}

\bibitem[{Beer and Williams(2015)}]{beer2015information}
Beer, R.~D. and Williams, P.~L. (2015).
\newblock Information processing and dynamics in minimally cognitive agents.
\newblock \emph{Cognitive science} 39, 1--38
\bibAnnoteFile{beer2015information}

\bibitem[{Clewley(2012)}]{clewley2012hybrid}
Clewley, R. (2012).
\newblock Hybrid models and biological model reduction with pydstool.
\newblock \emph{PLoS Computational Biology} 8, e1002628
\bibAnnoteFile{clewley2012hybrid}

\bibitem[{Dasgupta et~al.(2013)Dasgupta, W\"{o}rg\"{o}tter, and
  Manoonpong}]{dasgupta2013information}
Dasgupta, S., W\"{o}rg\"{o}tter, F., and Manoonpong, P. (2013).
\newblock Information dynamics based self-adaptive reservoir for delay temporal
  memory tasks.
\newblock \emph{Evolving Systems} 4, 235--249
\bibAnnoteFile{dasgupta2013information}

\bibitem[{de~Wit et~al.(2012)de~Wit, Siciliano, and Bastin}]{de2012theory}
de~Wit, C.~C., Siciliano, B., and Bastin, G. (2012).
\newblock \emph{Theory of robot control} (Springer Science \& Business Media)
\bibAnnoteFile{de2012theory}

\bibitem[{Der and Martius(2012)}]{der2012playful}
Der, R. and Martius, G. (2012).
\newblock \emph{The Playful Machine: Theoretical Foundation and Practical
  Realization of Self-Organizing Robots}, vol.~15 (Springer Science \& Business
  Media)
\bibAnnoteFile{der2012playful}

\bibitem[{Der and Martius(2015)}]{der2015novel}
Der, R. and Martius, G. (2015).
\newblock A novel plasticity rule can explain the development of sensorimotor
  intelligence.
\newblock \emph{arXiv preprint arXiv:1505.00835}
\bibAnnoteFile{der2015novel}

\bibitem[{Dorf and Bishop(1998)}]{dorf1998modern}
Dorf, R.~C. and Bishop, R.~H. (1998).
\newblock \emph{Modern control systems} (Pearson (Addison-Wesley))
\bibAnnoteFile{dorf1998modern}

\bibitem[{Echeveste et~al.(2015)Echeveste, Eckmann, and
  Gros}]{echeveste2015fisher}
Echeveste, R., Eckmann, S., and Gros, C. (2015).
\newblock The fisher information as a neural guiding principle for independent
  component analysis.
\newblock \emph{Entropy} 17, 3838--3856
\bibAnnoteFile{echeveste2015fisher}

\bibitem[{Echeveste and Gros(2014)}]{echeveste2014generating}
Echeveste, R. and Gros, C. (2014).
\newblock Generating functionals for computational intelligence: The fisher
  information as an objective function for self-limiting hebbian learning
  rules.
\newblock \emph{Frontiers in Robotics and AI} 1
\bibAnnoteFile{echeveste2014generating}

\bibitem[{Egbert et~al.(2010)Egbert, Barandiaran, and {Di
  Paolo}}]{egbert2010minimal}
Egbert, M.~D., Barandiaran, X.~E., and {Di Paolo}, E.~A. (2010).
\newblock A minimal model of metabolism-based chemotaxis.
\newblock \emph{PLoS computational biology} 6, e1001004
\bibAnnoteFile{egbert2010minimal}

\bibitem[{Ernesti et~al.(2012)Ernesti, Righetti, Do, Asfour, and
  Schaal}]{ernesti2012encoding}
Ernesti, J., Righetti, L., Do, M., Asfour, T., and Schaal, S. (2012).
\newblock Encoding of periodic and their transient motions by a single dynamic
  movement primitive.
\newblock In \emph{2012 12th IEEE-RAS International Conference on Humanoid
  Robots (Humanoids 2012)} (IEEE), 57--64
\bibAnnoteFile{ernesti2012encoding}

\bibitem[{Friston(2010)}]{friston2010free}
Friston, K. (2010).
\newblock The free-energy principle: a unified brain theory?
\newblock \emph{Nature Reviews Neuroscience} 11, 127--138
\bibAnnoteFile{friston2010free}

\bibitem[{Friston and Ao(2011)}]{friston2011free}
Friston, K. and Ao, P. (2011).
\newblock Free energy, value, and attractors.
\newblock \emph{Computational and mathematical methods in medicine} 2012
\bibAnnoteFile{friston2011free}

\bibitem[{Gros(2015)}]{gros2015complex}
Gros, C. (2015).
\newblock \emph{Complex and adaptive dynamical systems: A primer} (Springer)
\bibAnnoteFile{gros2015complex}

\bibitem[{Gros et~al.(2014)Gros, Linkerhand, and Walther}]{gros2014attractor}
Gros, C., Linkerhand, M., and Walther, V. (2014).
\newblock Attractor metadynamics in adapting neural networks.
\newblock In \emph{Artificial Neural Networks and Machine Learning--ICANN 2014}
  (Springer). 65--72
\bibAnnoteFile{gros2014attractor}

\bibitem[{Hobbelen(2008)}]{hobbelen2008limit}
Hobbelen, D.~G. (2008).
\newblock \emph{Limit cycle walking} (TU Delft, Delft University of Technology)
\bibAnnoteFile{hobbelen2008limit}

\bibitem[{Ijspeert(2008)}]{ijspeert2008central}
Ijspeert, A.~J. (2008).
\newblock Central pattern generators for locomotion control in animals and
  robots: a review.
\newblock \emph{Neural Networks} 21, 642--653
\bibAnnoteFile{ijspeert2008central}

\bibitem[{Ijspeert et~al.(2013)Ijspeert, Nakanishi, Hoffmann, Pastor, and
  Schaal}]{ijspeert2013dynamical}
Ijspeert, A.~J., Nakanishi, J., Hoffmann, H., Pastor, P., and Schaal, S.
  (2013).
\newblock Dynamical movement primitives: learning attractor models for motor
  behaviors.
\newblock \emph{Neural computation} 25, 328--373
\bibAnnoteFile{ijspeert2013dynamical}

\bibitem[{Ijspeert et~al.(2002)Ijspeert, Nakanishi, and
  Schaal}]{ijspeert2002learning}
Ijspeert, A.~J., Nakanishi, J., and Schaal, S. (2002).
\newblock Learning attractor landscapes for learning motor primitives.
\newblock In \emph{Advances in Neural Information Processing Systems} (MIT
  Press), 1547--1554
\bibAnnoteFile{ijspeert2002learning}

\bibitem[{Kelso(1994)}]{kelso1994informational}
Kelso, J. (1994).
\newblock The informational character of self-organized coordination dynamics.
\newblock \emph{Human Movement Science} 13, 393--413
\bibAnnoteFile{kelso1994informational}

\bibitem[{Klyubin et~al.(2005)Klyubin, Polani, and
  Nehaniv}]{klyubin2005empowerment}
Klyubin, A.~S., Polani, D., and Nehaniv, C.~L. (2005).
\newblock Empowerment: A universal agent-centric measure of control.
\newblock In \emph{Evolutionary Computation, 2005. The 2005 IEEE Congress on}
  (IEEE), vol.~1, 128--135
\bibAnnoteFile{klyubin2005empowerment}

\bibitem[{Laszlo et~al.(1996)Laszlo, van~de Panne, and Fiume}]{laszlo1996limit}
Laszlo, J., van~de Panne, M., and Fiume, E. (1996).
\newblock Limit cycle control and its application to the animation of balancing
  and walking.
\newblock In \emph{Proceedings of the 23rd annual conference on Computer
  graphics and interactive techniques} (ACM), 155--162
\bibAnnoteFile{laszlo1996limit}

\bibitem[{Lizier et~al.(2012)Lizier, Prokopenko, and Zomaya}]{lizier2012local}
Lizier, J.~T., Prokopenko, M., and Zomaya, A.~Y. (2012).
\newblock Local measures of information storage in complex distributed
  computation.
\newblock \emph{Information Sciences} 208, 39--54
\bibAnnoteFile{lizier2012local}

\bibitem[{Markovi{\'c} and Gros(2010)}]{markovic2010self}
Markovi{\'c}, D. and Gros, C. (2010).
\newblock Self-organized chaos through polyhomeostatic optimization.
\newblock \emph{Physical Review Letters} 105, 068702
\bibAnnoteFile{markovic2010self}

\bibitem[{Markovi{\'c} and Gros(2012)}]{markovic2012intrinsic}
Markovi{\'c}, D. and Gros, C. (2012).
\newblock Intrinsic adaptation in autonomous recurrent neural networks.
\newblock \emph{Neural Computation} 24, 523--540
\bibAnnoteFile{markovic2012intrinsic}

\bibitem[{Martius et~al.(2013)Martius, Der, and Ay}]{martius2013information}
Martius, G., Der, R., and Ay, N. (2013).
\newblock Information driven self-organization of complex robotic behaviors.
\newblock \emph{PLOS ONE} 8, e63400
\bibAnnoteFile{martius2013information}

\bibitem[{Nolfi and Floreano(2000)}]{nolfi2000evolutionary}
Nolfi, S. and Floreano, D. (2000).
\newblock \emph{Evolutionary Robotics: The Biology, Intelligence, and
  Technology of Self-organizing Machines} (MIT Press)
\bibAnnoteFile{nolfi2000evolutionary}

\bibitem[{Olsson et~al.(2006)Olsson, Nehaniv, and Polani}]{olsson2006unknown}
Olsson, L.~A., Nehaniv, C.~L., and Polani, D. (2006).
\newblock From unknown sensors and actuators to actions grounded in
  sensorimotor perceptions.
\newblock \emph{Connection Science} 18, 121--144
\bibAnnoteFile{olsson2006unknown}

\bibitem[{Pfeifer et~al.(2007)Pfeifer, Lungarella, and Iida}]{pfeifer2007self}
Pfeifer, R., Lungarella, M., and Iida, F. (2007).
\newblock Self-organization, embodiment, and biologically inspired robotics.
\newblock \emph{Science} 318, 1088--1093
\bibAnnoteFile{pfeifer2007self}

\bibitem[{S\'{a}ndor and Gros(2015)}]{sandor2015general}
S\'{a}ndor, B. and Gros, C. (2015).
\newblock A versatile class of prototype dynamical systems for complex
  bifurcation cascades of limit cycles.
\newblock \emph{Scientific Reports} 5, 12316
\bibAnnoteFile{sandor2015general}

\bibitem[{Schaal et~al.(2000)Schaal, Kotosaka, and
  Sternad}]{schaal2000nonlinear}
Schaal, S., Kotosaka, S., and Sternad, D. (2000).
\newblock Nonlinear dynamical systems as movement primitives.
\newblock In \emph{IEEE International Conference on Humanoid Robotics}. 1--11
\bibAnnoteFile{schaal2000nonlinear}

\bibitem[{Schmidt et~al.(2013)Schmidt, Hoffmann, Nakajima, and
  Pfeifer}]{schmidt2013bootstrapping}
Schmidt, N.~M., Hoffmann, M., Nakajima, K., and Pfeifer, R. (2013).
\newblock Bootstrapping perception using information theory: case studies in a
  quadruped robot running on different grounds.
\newblock \emph{Advances in Complex Systems} 16, 1250078
\bibAnnoteFile{schmidt2013bootstrapping}

\bibitem[{Taga et~al.(1991)Taga, Yamaguchi, and Shimizu}]{taga1991self}
Taga, G., Yamaguchi, Y., and Shimizu, H. (1991).
\newblock Self-organized control of bipedal locomotion by neural oscillators in
  unpredictable environment.
\newblock \emph{Biological cybernetics} 65, 147--159
\bibAnnoteFile{taga1991self}

\bibitem[{Tani and Ito(2003)}]{tani2003self}
Tani, J. and Ito, M. (2003).
\newblock Self-organization of behavioral primitives as multiple attractor
  dynamics: A robot experiment.
\newblock \emph{Systems, Man and Cybernetics, Part A: Systems and Humans, IEEE
  Transactions on} 33, 481--488
\bibAnnoteFile{tani2003self}

\bibitem[{Toyoizumi et~al.(2005)Toyoizumi, Pfister, Aihara, and
  Gerstner}]{toyoizumi2005generalized}
Toyoizumi, T., Pfister, J.-P., Aihara, K., and Gerstner, W. (2005).
\newblock Generalized bienenstock-cooper-munro rule for spiking neurons that
  maximizes information transmission.
\newblock \emph{Proceedings of the National Academy of Sciences of the United
  States of America} 102, 5239--44
\bibAnnoteFile{toyoizumi2005generalized}

\bibitem[{Triesch(2005)}]{triesch2005gradient}
Triesch, J. (2005).
\newblock A gradient rule for the plasticity of a neuron’s intrinsic
  excitability.
\newblock In \emph{Artificial Neural Networks: Biological Inspirations–ICANN
  2005} (Springer). 65--70
\bibAnnoteFile{triesch2005gradient}

\bibitem[{Triesch(2007)}]{triesch2007synergies}
Triesch, J. (2007).
\newblock Synergies between intrinsic and synaptic plasticity mechanisms.
\newblock \emph{Neural Computation} 19, 885--909
\bibAnnoteFile{triesch2007synergies}

\bibitem[{Williams and Beer(2013)}]{williams2013environmental}
Williams, P. and Beer, R. (2013).
\newblock Environmental feedback drives multiple behaviors from the same neural
  circuit.
\newblock In \emph{Advances in Artificial Life, ECAL 2013} (MIT Press),
  vol.~12, 268--275
\bibAnnoteFile{williams2013environmental}

\bibitem[{Ziemke(2003)}]{ziemke2003s}
Ziemke, T. (2003).
\newblock What’s that thing called embodiment.
\newblock In \emph{Proceedings of the 25th Annual meeting of the Cognitive
  Science Society} (Mahwah, NJ: Lawrence Erlbaum), 1305--1310
\bibAnnoteFile{ziemke2003s}

\end{thebibliography}

\end{document}